\documentclass{kapedbk}
\usepackage{graphicx,amsmath,amsfonts,edbkps}
\normallatexbib

\newcommand{\ket}[1]{{| #1 \rangle}}
\newcommand{\up}{\!\uparrow}
\newcommand{\down}{\!\downarrow}
\newcommand{\eps}{\epsilon}

\newcommand{\second}{2^{\mathrm{nd}}}
\newcommand{\then}{\Rightarrow }

\begin{document}

\articletitle{Creation and Detection of mobile and non-local spin-entangled electrons}

\author{Patrik Recher\thanks{Present addresses: E.L. Ginzton Laboratory, Stanford University, Stanford, California 94305, USA, and Institute of 
Industrial Science, University of Tokyo, 4-6-1 Komaba, Meguro-ku, Tokyo 153-8505, Japan}, D.S. Saraga, and Daniel Loss}
\affil{Department of Physics and Astronomy, University of
Basel,\\
Klingelbergstrasse 82, CH-4056 Basel, Switzerland}

\begin{abstract}
We present electron spin entanglers--devices creating mobile spin-entangled electrons that are spatially separated--where the spin-entanglement 
in a superconductor present in form of Cooper pairs and in a single quantum dot with a spin singlet groundstate is transported to two spatially 
separated leads by means of a correlated two-particle tunneling event. The unwanted process of both electrons tunneling into the same lead is 
suppressed by strong Coulomb blockade effects caused by quantum dots, Luttinger liquid effects or by resistive outgoing leads. In this review we 
give  a transparent description of the different setups, including discussions of the feasibility of the subsequent detection of 
spin-entanglement via charge noise measurements. Finally, we show that quantum dots in the spin filter regime can be used to perform Bell-type 
measurements that only require the measurement of zero frequency charge noise correlators.
\end{abstract}

\begin{keywords}
Entanglement, Andreev tunneling, quantum dots, Luttinger liquids, Coulomb blockade, Bell inequalities, spin filtering
\end{keywords}


\section{Sources of mobile spin-entangled electrons}
\label{entanglers}

\footnote{These proceedings are published in "Fundamental Problems of Mesoscopic Physics Interaction and Decoherence", pp. 179-202, eds. I.V. 
Lerner et al., NATO Science Ser. II, Vol. 154 (Kluwer, Dordrecht, 2004).} The extensive search for mechanisms to create electronic entanglement 
in solid state systems was motivated partly by the idea to use spin \cite{Loss97} or charge \cite{Barenco} degrees of freedom of electrons in 
quantum confined nanostructures as a quantum bit (qubit) for quantum computing. In particular, pairwise entangled states are the basic 
ingredients to perform elementary quantum gates \cite{Loss97}. Furthermore, exploiting the charge of electrons allows to easily transport such 
entangled states along wires by means of electric fields, leading to mobile and non-local entangled states. These are required for quantum 
communication protocols as well as in experiments where nonlocality and entanglement are detected via the violation of a Bell inequality suitably 
formulated for massive particles in a solid state environment. We will turn to this issue in Section~\ref{Detection}.
One should 
note that entanglement is rather the rule than the exception in
solid state systems, as it arises naturally from Fermi statistics. For
instance, the ground state of a helium atom is the spin singlet
$|\!\!\uparrow \downarrow \rangle - |\!\!\downarrow \uparrow
\rangle$. Similarly, one finds a singlet in the ground state of a
quantum dot with two electrons \cite{Pfannkuche}. However,
such ``local'' entangled singlets are not readily useful for
quantum computation and communication, as these require control
over each individual electron as well as non-local correlations.
An improvement in this direction is given by two coupled quantum
dots with a single electron in each dot \cite{Loss97}, where the
spin-entangled electrons are already spatially separated by strong
on-site Coulomb repulsion (like in a hydrogen molecule). One could then create mobile entangled electrons by
simultaneously lowering the tunnel barriers coupling each dot to
separate leads. Another natural source of spin entanglement can be
found in superconductors, as these contain Cooper pairs with
singlet spin wave functions. It was first shown in Ref.
\cite{P1CBL} how a non-local entangled state is created in two
uncoupled quantum dots when they are coupled to the same superconductor. In
a non-equilibrium situation, the Cooper pairs can be extracted to
normal leads by Andreev tunneling, thus creating a flow of
entangled pairs \cite{P3RSL,P16Lesovik,P8RL,P13BVBF,Samuelsson,RL2}.

A crucial requirement for an entangler is to create {\em spatially
separated} entangled electrons; hence one must avoid whole
entangled pairs entering the same lead. As will be shown below,
energy conservation is an efficient mechanism for the suppression
of undesired channels. For this, interactions can play a decisive
role. For instance, one can use Coulomb repulsion in quantum dots
\cite{P3RSL},\cite{Sar02}, in Luttinger liquids \cite{P8RL},\cite{P13BVBF} or in a setup where resistive leads give rise to a dynamical Coulomb 
blockade  effect \cite{RL2}. Finally, we mention other entangler
proposals using leads with narrow bandwidth \cite{Oliver}
and/or generic quantum interference effects \cite{Bos02,Saraga2}. 

In the
following sections we present  our theoretical proposals towards the implementation of a
solid-state
entangler.

\section{Superconductor-based electron spin-entanglers}
\label{ssentanglers}

Here we envision a  {\em non-equilibrium} situation in which the
electrons of a Cooper pair can tunnel coherently by means of an
Andreev tunneling event from a superconductor to two separate normal leads,
one electron per lead. Due to an applied bias voltage, the
electron pairs can move into the leads thus giving rise to mobile
spin entanglement. Note that an (unentangled) single-particle
current is strongly suppressed by energy conservation as long as
both the temperature and the bias are  much smaller than the
superconducting gap. In the following we review three proposals
where we exploit the repulsive Coulomb charging energy between the
two spin-entangled electrons in order to separate them so that the
residual current in the leads is carried by non-local singlets. We
show that such entanglers meet all requirements necessary for subsequent
detection of spin-entangled electrons via charge noise measurements discussed in Section~\ref{Detection}.


\subsection{Andreev Entangler with quantum dots}
\label{ssecAndreev}

\begin{figure}[h]
\centerline{\includegraphics[width=5.5cm]{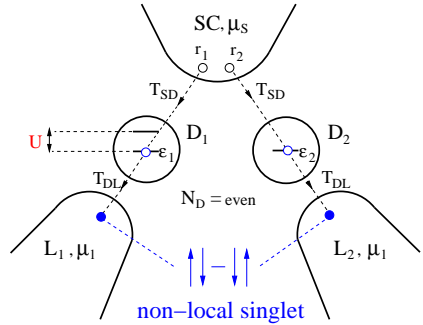}}
\vspace{5mm}
\caption{The entangler setup. Two spin-entangled electrons forming
a Cooper pair tunnel with amplitude $T_{SD}$ from points ${\bf
r}_{1}$ and ${\bf r}_{2}$ of the superconductor, SC, to two dots,
$D_{1}$ and $D_{2}$, by means of Andreev tunneling. The dots are
tunnel-coupled to normal Fermi liquid leads ${\rm L_{1}}$ and
${\rm L_{2}}$, with tunneling amplitude $T_{DL}$. The
superconductor and leads are kept at chemical potentials $\mu_{S}$
and $\mu_{l}$, respectively. Adapted from \cite{P3RSL}.}
\label{spintabl}
\end{figure}

The proposed entangler setup (see Fig.~\ref{spintabl}) consists of
a superconductor (SC) with chemical potential $\mu_{S}$ which is weakly coupled to
two quantum dots (QDs) in the Coulomb blockade regime
\cite{Kou97}. These QDs are in turn  weakly coupled to outgoing
Fermi liquid leads, held at the same chemical potential $\mu_{l}$. Note that in the presence of a voltage bias between the two leads 1,2 an 
(unentangled) current could  flow from one lead to the other via the SC.
A bias voltage $\mu=\mu_{S}-\mu_{l}$ is applied between the
SC and the leads. The tunneling amplitudes between the SC and the
dots, and dots and leads, are denoted by $T_{SD}$ and $T_{DL}$,
respectively (see Fig.~\ref{spintabl}).
The two intermediate QDs in the Coulomb blockade regime have
chemical potentials $\epsilon_{1}$ and $\epsilon_{2}$,
respectively. These can be tuned via external gate voltages, such
that the tunneling of two electrons via different dots into
different leads is resonant for
$\epsilon_{1}+\epsilon_{2}=2\mu_{S}$ \cite{energyconservation}. As
it turns out \cite{P3RSL}, this two-particle resonance is
suppressed for the tunneling of two electrons via the same dot
into the same lead by the on-site repulsion $U$ of the dots and/or
the superconducting gap $\Delta$. Next, we specify the parameter
regime of interest here in which the initial spin-entanglement of
a Cooper pair in the SC is successfully transported to the leads.

Besides the fact that single-electron tunneling and tunneling of
two electrons via the same dot should be excluded, we also have to
suppress transport of electrons which are already on the QDs.
This could lead to effective spin-flips on the QDs, which would
destroy the spin entanglement of the two electrons tunneling into
the Fermi leads. A further source of unwanted spin-flips on the
QDs is provided by its coupling to the Fermi liquid leads via
particle-hole excitations in the leads. The QDs can be treated
each as one localized spin-degenerate level as long as the mean
level spacing $\delta\epsilon$ of the dots exceeds both the bias
voltage $\mu$ and the temperature $k_{B}T$. In addition, we
require that each QD contains an even number of electrons with a
spin-singlet ground state. A more detailed analysis of such a
parameter regime is given in \cite{P3RSL} and is stated here
\begin{equation}
\label{regime} \Delta,U,
\delta\epsilon>\mu>\gamma_{l},k_{B}T,{\rm
and}\,\,\gamma_{l}>\gamma_{S}.
\end{equation}
In Eq.~(\ref{regime}) the rates for tunneling of an electron from the
SC to the QDs and from the QDs to the Fermi leads are given by
$\gamma_{S}=2\pi\nu_{S}|T_{SD}|^{2}$  and
$\gamma_{l}=2\pi\nu_{l}|T_{DL}|^{2}$, respectively,  with
$\nu_{S}$ and $\nu_{l}$ being the corresponding  electron density
of states per spin at the Fermi level. We consider asymmetric
barriers $\gamma_{l}>\gamma_{s}$ in order to exclude correlations
between subsequent Cooper pairs on the QDs. We work at the
particular interesting resonance $\epsilon_{1},\epsilon_{2}\simeq
\mu_{S}$, where the injection of the electrons into different
leads takes place at the same orbital energy. This is a crucial
requirement for the subsequent detection of entanglement via noise
\cite{beamsplitter}. In this regime, we have calculated and
compared the stationary charge current of two spin-entangled
electrons for two competing transport channels in a T-matrix approach \cite{Merzbacher}.

As a result, the ratio of the desired current for two electrons tunneling into
{\em different} leads ($I_{1}$) to the unwanted current for two
electrons into the {\em same} lead ($I_{2}$) is~\cite{P3RSL}
\begin{equation}
\label{final} \frac{I_{1}}{I_{2}}= \frac{4{\cal E}^2}{\gamma^2}
\left[\frac{\sin(k_{F}\delta r)}{k_{F}\delta
r}\right]^2\,e^{-2\delta r/\pi\xi}, \quad\quad\quad
   \frac{1}{{\cal E}}=\frac{1}{\pi\Delta}+\frac{1}{U},
\end{equation}
where $\gamma = \gamma_1 + \gamma_2$. The current $I_1=(4e\gamma_{S}^{2}/\gamma)(\sin(k_{F}\delta r)/k_{F}\delta r)^{2}e^{-2\delta r/\pi\xi}$ 
becomes
exponentially suppressed with increasing distance $\delta r=|{\bf
r}_1-{\bf r}_2|$ between the tunneling points on the SC, on a
scale given by the superconducting coherence length $\xi$ which determines
the size of a Cooper pair. This does not pose a severe restriction
for conventional s-wave materials  with $\xi$ typically being on
the order of $\mu {\rm m}$. In the relevant case $\delta r<\xi$
the suppression is only polynomial $\propto 1/(k_{F}\delta r)^2$,
with $k_{F}$ being the Fermi wave number in the SC. On the other
hand, we see that the effect of the QDs  consists in the
suppression factor $(\gamma/{\cal E})^2$ for tunneling into the
same lead \cite{cost}. Thus, in addition to Eq.~(\ref{regime}) we
have to impose the condition $k_F\delta r < {\cal E}/\gamma$,
which is  well satisfied for small dots with ${\cal E}/\gamma\sim
100$ and for $\delta r\sim$ 1nm. As an experimental probe to test if
the two spin-entangled electrons indeed separate and tunnel to
different leads we suggest to join the two leads 1 and 2 to form
an Aharonov-Bohm loop. In such a setup the different tunneling
paths of an Andreev process from the SC via the dots to the leads
can interfere. As a result, the measured current as a function of
the applied magnetic flux $\phi$ threading the loop contains a
phase coherent part $I_{AB}$ which consists of oscillations with
periods $h/e$ and $h/2e$ \cite{P3RSL}
\begin{equation}
\label{AB-osc} I_{AB}\sim
\sqrt{8I_{1}I_{2}}\cos(\phi/\phi_{0})+I_{2}\cos(2\phi/\phi_{0}),
\end{equation}
with $\phi_{0}=h/e$ being the single-electron flux quantum. The
ratio of the two contributions scales like $\sqrt{I_{1}/I_{2}}$
which suggest that by decreasing $I_{2}$ (e.g. by increasing $U$)
the $h/2e$ oscillations should vanish faster than the $h/e$ ones.

We note that the efficiency as well as the absolute rate for the
desired injection of two electrons into different  leads can 
be enhanced by using lower dimensional SCs \cite{P8RL} . In
two dimensions (2D) we find that $I_{1}\propto 1/k_{F}\delta r$
for large $k_{F}\delta r$, and in one dimension (1D) there is no
suppression of the current and only an oscillatory behavior in
$k_{F}\delta r $ is found. A 2D-SC can be realized by using a SC
on top of a two-dimensional electron gas (2DEG) \cite{P4Klapwijk,Weiss},
where superconducting correlations are induced via the proximity
effect in the 2DEG. In 1D, superconductivity was found in ropes of
single-walled carbon nanotubes \cite{P5Bouchiat}.

Finally, we note that the coherent injection of Cooper pairs by an
Andreev process allows the detection of individual spin-entangled
electron pairs in the leads. The delay time $\tau_{\rm delay}$
between the two electrons of a pair is given by $1/\Delta$,
whereas the separation in time of subsequent pairs is given
approximately by $\tau_{\rm pairs} \sim 2e/I_{1}\sim
\gamma_{l}/\gamma_{S}^{2}$ (up to geometrical factors). For $\gamma_{S}\sim\gamma_{l}/10\sim 1\mu {\rm eV}$
and $\Delta\sim 1{\rm meV}$ we obtain that the delay time
$\tau_{\rm delay}\sim 1/\Delta\sim 1{\rm ps}$ is much smaller than
the average delivery time $\tau_{\rm pairs}$ per entangled pair
$2e/I_{1}\sim 40{\rm ns}$. Such a time separation is indeed
necessary in order to detect individual pairs of spin-entangled
electrons. We return to this issue in Section~\ref{Detection}.

\subsection{Andreev Entangler with Luttinger liquid leads}
\label{sssluttinger}

\begin{figure}[h]
\centerline{\includegraphics[width=5.5cm]{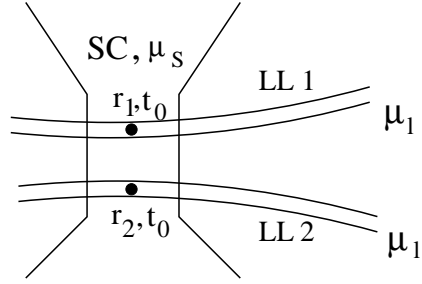}}
\vspace{5mm} 
\caption{Two quantum wires 1,2, with chemical
potential $\mu_{l}$ and described as infinitely long Luttinger
liquids (LLs), are deposited on top of an s-wave superconductor
(SC) with chemical potential $\mu_{S}$. The electrons of a Cooper
pair can tunnel by means of an Andreev process from two points
${\bf r}_{1}$ and ${\bf r}_{2}$ on the SC to the center (bulk) of
the two quantum wires 1 and 2, respectively, with tunneling
amplitude $t_{0}$. Adapted from \cite{P8RL}.}\label{LLfig}
\end{figure}

Next, we discuss a setup with an s-wave SC weakly coupled  to the
center (bulk) of two separate one-dimensional leads (quantum
wires) 1,2 (see Fig.~\ref{LLfig}) which exhibit Luttinger liquid
(LL) behavior, such as carbon nanotubes
\cite{P6Bockrath,P10Egger,P11Kane} or in semiconducting cleaved edge quantum wires \cite{Auslaender}. The leads are assumed to be
infinitely extended and are described by conventional LL-theory
\cite{P7rev}.

Interacting electrons in one dimension lack the existence of quasi
particles like they exist in a Fermi liquid and instead the low
energy excitations are collective charge and spin modes. In the
absence of backscattering interaction the velocities of the charge
and spin excitations are given by $u_{\rho}=v_{F}/K_{\rho}$ for
the charge and $u_{\sigma}=v_{F}$ for the spin, where $v_{F}$ is
the Fermi velocity and $K_{\rho}<1$ for repulsive interaction
between electrons ($K_{\rho}=1$ corresponds to a 1D-Fermi gas). As
a consequence of this non-Fermi liquid behavior, tunneling into a
LL is strongly suppressed at low energies. Therefore one should
expect additional interaction effects in a coherent two-particle tunneling
event (Andreev process) of a Cooper pair from the SC to the leads.
We find that strong LL-correlations result in an additional
suppression for tunneling of two coherent electrons into the {\em
same} LL compared to single electron tunneling into a LL if the
applied bias voltage $\mu$ between the SC and the two leads is
much smaller than the energy gap $\Delta$ of the SC.

To quantify the effectiveness of such an entangler, we calculate
the current for the two competing processes of tunneling into
different leads ($I_{1}$) and into the same lead ($I_{2}$) in
lowest order via a tunneling Hamiltonian approach. Again, we
account for a finite distance separation $\delta r$ between the
two exit points on the SC when the two electrons of a Cooper pair
tunnel to different leads. For the current $I_{1}$ of the desired
pair-split process we obtain, in leading order in $\mu/\Delta$ and
at zero temperature~\cite{P8RL}
\begin{equation}
\label{LL1}
I_{1}=\frac{I_{1}^{0}}{\Gamma(2\gamma_{\rho}+2)}\frac{v_{F}}{u_{\rho}}
\left(\frac{2\mu\Lambda}{u_{\rho}}\right)^{2\gamma_{\rho}},
\,\,I_{1}^{0}=\pi e\gamma^{2}\mu F_{d}^{2}(\delta r),
\end{equation}
where $\Gamma (x)$ is the Gamma function and $\Lambda$ is a short
distance cut-off on the order of the lattice spacing in the LL and
$\gamma=4\pi\nu_{S}\nu_{l}|t_{0}|^{2}$ is the dimensionless tunnel
conductance per spin with $t_{0}$ being the bare tunneling
amplitude for electrons to tunnel from the SC to the LL-leads (see
Fig.~\ref{LLfig}). The electron density of states per spin at the
Fermi level for the SC and the LL-leads are denoted by $\nu_{S}$
and $\nu_{l}$, respectively. The current $I_{1}$ has its
characteristic non-linear form $I_{1}\propto
\mu^{2\gamma_{\rho}+1}$ with
$\gamma_{\rho}=(K_{\rho}+K_{\rho}^{-1})/4-1/2>0$ being the
exponent for tunneling into the bulk of a {\em single} LL. The
factor $F_{d}(\delta r)$ in Eq.~(\ref{LL1}) depends on the geometry of
the device and is given here again by $F_{d}(\delta
r)=[\sin(k_{F}\delta r)/k_{F}\delta r]\exp(-\delta r/\pi\xi)$
for the case of a 3D-SC. In complete analogy to
Section~\ref{ssecAndreev} the power law suppression in
$k_{F}\delta r$ is weaker for  lower dimensions of the SC.

This result should be compared with the unwanted transport channel
where two electrons of a Cooper pair tunnel into the same lead 1
or 2 but with $\delta r=0$.  We find that such processes are
indeed suppressed by strong LL-correlations if $\mu<\Delta$. The
result for the current ratio $I_{2}/I_{1}$ in leading order in
$\mu/\Delta$ and for zero temperature is \cite{P8RL}
\begin{equation}
\frac{I_{2}}{I_{1}}=F_{d}^{-1}(\delta r)\sum\limits_{b=\pm
1}\,A_{b}\,\left(\frac{2\mu}{\Delta}\right)^{2\gamma_{\rho
b}},\,\,\gamma_{\rho +}=\gamma_{\rho},\,\,\gamma_{\rho
-}=\gamma_{\rho}+(1-K_{\rho})/2, \label{currentI222}
\end{equation}
where $A_{b}$ is an interaction dependent
constant~\cite{LLfootnote}. The result (\ref{currentI222}) shows
that the current $I_{2}$ for injection of two electrons into the
same lead is suppressed compared to $I_{1}$ by a factor of
$(2\mu/\Delta)^{2\gamma_{\rho +}}$, if both electrons are injected
into the  same branch (left or right movers), or by
$(2\mu/\Delta)^{2\gamma_{\rho -}}$ if the two electrons travel in
different directions \cite{P21electronbunching}. The suppression
of the current $I_{2}$ by $1/\Delta$ reflects the two-particle
correlation effect in the LL, when the electrons tunnel into the
same lead. The larger $\Delta$, the shorter the delay time is
between the arrivals of the two partner electrons of a Cooper
pair, and, in turn, the more the second electron tunneling into
the same lead will feel the existence of the first one which is
already present in the LL. This behavior is similar to the Coulomb
blockade effect in QDs, see Section~\ref{ssecAndreev}. Concrete
realizations of LL-behavior is found in metallic carbon nanotubes
with similar exponents as derived here \cite{P10Egger,P11Kane}. In
metallic single-walled carbon nanotubes $K_{\rho}\sim 0.2$
\cite{P6Bockrath} which corresponds to $2\gamma_{\rho}\sim  1.6$.
This suggests the rough estimate $(2\mu/\Delta)<1/k_{F}\delta r$
for the entangler to be efficient. As a consequence, voltages in
the range $k_{B}T<\mu<100 \mu {\rm eV}$ are required for $\delta
r\sim 1$ nm and $\Delta\sim 1{\rm meV}$.  In addition, nanotubes
were reported to be very good spin conductors \cite{P12Balents}
with estimated spin-flip scattering lengths of the order of $\mu
{\rm m}$ \cite{P13BVBF}. In GaAs quantum wires $K_{\rho}\sim 0.66-0.82$ \cite{Auslaender} which suggests that interaction is also pronounced in 
such systems.

We now briefly address the question of spin and charge transport in a LL. Let's suppose that two electrons of a pair tunnel to different LLs 
(desired pair-split process). We assume that an electron with spin $s=\pm 1/2$ tunnels into a given lead (1 or 2) as a right-mover and at point 
$x$. We then create the state $|\alpha\rangle=\psi_{s}^{\dagger}(x)|0\rangle$ where $|0\rangle$ denotes the ground state of the LL. We now 
consider the time evolution of the charge density fluctuations $\rho(x')=\sum_{s}:\,\psi_{s}^{\dagger}(x')\psi_{s}(x')\,: $ and the spin density 
fluctuations $\sigma^{z}(x)=\sum_{s}s:\,\psi_{s}^{\dagger}(x')\psi_{s}(x')\,: $ where $:\,\,\,:$ denotes normal ordering. We then obtain for the 
charge propagation
\begin{equation}
\langle\alpha|\rho(x',t)|\alpha\rangle=\frac{1}{2}(1+K_{\rho})\delta(x'-x-u_{\rho}t)+\frac{1}{2}(1-K_{\rho})\delta(x'+x+u_{\rho}t)
\end{equation}
and for the spin propagation
\begin{equation}
\langle\alpha|\sigma^{z}(x',t)|\alpha\rangle=s\delta(x'-x-u_{\sigma}t).
\end{equation}
The shape of the $\delta-$function is unchanged with time due to the linear spectrum of the LL model. In reality, carbon nanotubes show such a 
highly linear spectrum up to energies of $\sim 1$ eV. Therefore, we expect that the injected spin is locally accessible  in carbon nanotubes but 
carried by the collective spin modes rather than by a single electron. Another interesting feature characteristic for a LL is the different 
propagation velocities for the charge and for the spin ($u_{\sigma}\neq u_{\rho}$) which is known as spin-charge separation.

\subsection{Andreev Entangler with resistive leads}
\label{resistiveleads}

Here we consider resistive normal leads weakly coupled to the SC. This gives rise to a dynamical Coulomb blockade (CB) effect with the 
consequence that in a pair tunneling process into the same lead the second electron still experiences the Coulomb repulsion of the first one, 
which has not yet diffused away. Such a setup is presumably simpler to realize experimentally than the setups introduced above. Natural existing 
candidates for such a setup with long spin decoherence lengths ($\sim 100$ $\mu$m \cite{Awschalom}) are semiconductor systems tunnel-coupled to a 
SC, as experimentally implemented in InAs \cite{InAs}, InGaAs \cite{Franceschi} or GaAs/AlGaAs \cite{Marsh}. Recently, 2DEGs with a resistance 
per square approaching the quantum resistance $R_{Q}=h/e^{2}\sim 25.8$ k$\Omega$ could be achieved by depleting the 2DEG with a voltage applied 
between a back gate and the 2DEG \cite{Rimberg}. In metallic normal NiCr leads of width $\sim 100$ nm and length $\sim 10$ $\mu$m, resistances of 
$R=22-24$ k$\Omega$ have been produced at low temperatures. Even larger resistances $R=200-250$ k$\Omega$ have been measured in Cr leads 
\cite{Kuzmin}.

\begin{figure}[ht]
\vskip.2in
\centerline{\includegraphics[width=5cm]{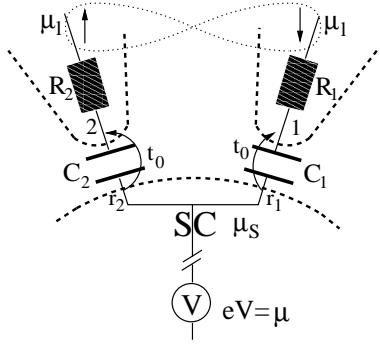}}
\caption{Entangler setup: A BCS bulk superconductor (SC) with chemical potential $\mu_{S}$ is tunnel-coupled (amplitude $t_{0}$)  via two points 
${\bf r}_{1}$ and ${\bf r}_{2}$ of the SC to two Fermi liquid leads 1,2 with resistance $R_{1,2}$. The two leads are held at the same chemical 
potential $\mu_{l}$ such that a  bias voltage $\mu=\mu_{S}-\mu_{l}$ is applied between the SC and the two leads via the voltage source V. The 
tunnel-junctions 1,2 have capacitances $C_{1,2}$. Adapted from \cite{RL2}.} 
\label{Resistancefig}
\end{figure}

The SC is held at the (electro-)chemical potential $\mu_{S}$  by a voltage source V, see Fig.~\ref{Resistancefig}. The two electrons of a Cooper 
pair can tunnel via two junctions placed at points ${\bf r}_{1}$ and ${\bf r}_{2}$ on the SC to two separate normal leads 1 and 2 with 
resistances $R_{1}$ and $R_{2}$, {\it resp.} They are kept at the same chemical potential $\mu_{l}$ so that a bias voltage $\mu=\mu_{S}-\mu_{l}$ 
is applied between SC and leads.
The system Hamiltonian decomposes into three parts $
H=H^{e}+H_{env}+H_{T}$. Here $H^{e}=H_{S}+\sum_{n=1,2}H_{ln}$ describes the
electronic parts of the isolated subsystems consisting of the SC and Fermi liquid leads $n=1,2$.

To describe resistance and dissipation in the normal leads we use a phenomenological approach \cite{Devoret}, where the electromagnetic 
fluctuations  in the circuit (being bosonic excitations) due to electron-electron interaction and the lead resistances are modeled by a bath of 
harmonic oscillators which is linearly coupled to the charge fluctuation ${Q_{n}}$ of the junction capacitor $n$ (induced by the tunneling 
electron). This physics is described by \cite{Devoret,oscillators}
\begin{equation}
\label{env}
 H_{env, n}=
\frac{{ Q_{n}}^{2}}{2C_{n}}+\sum\limits_{j=1}^{N}\,\left[\frac{q_{nj}^{2}}{2C_{nj}}+\frac{({\phi_{n}}-\varphi_{nj})^{2}}{2e^{2}L_{nj}}\right].
\end{equation}
The phase ${\phi_{n}}$ of junction $n$ is the conjugate variable to the charge satisfying $[{\phi_{n}},{Q_{m}}]=ie\delta_{n,m}$.
As a consequence $e^{-i{\phi_{n}}}$ reduces 
${Q_{n}}$ by one elemantary charge $e$.
As long as the cross capacitance $C_{12}$ between the two leads 1 and 2 is much smaller than the junction capacitances $C_{1,2}$ the charge 
relaxations of both tunnel junctions occur  independently of each other. As a result we have $H_{env}=\sum_{n=1,2}H_{env,n}$. The tunnel 
Hamiltonian $H_{T}$ now contains an additional phase factor due to the coupling of the tunneling electron to the environment, i.e. $ 
H_{T}=t_{0}\sum_{n,\sigma}\,\psi_{n\sigma}^{\dagger}\Psi_{\sigma}({\bf
r}_{n})\,e^{-i{\phi}_{n}}+{\rm h.c.}$ This phase factor obeys the following correlation function 
$\langle\exp(i{\phi_{n}}(t))\exp(-i{\phi_{n}(0)})\rangle=\exp[J(t)]$ with $J(t)=2\int_{0}^{\infty}(d\omega/\omega) ({\rm Re} 
Z_{T}(\omega)/R_{Q})(\exp(-i\omega t)-1)$. Here we introduced the total impedance $Z_{T}=(i\omega C+R^{-1})^{-1}$, with a purely Ohmic lead 
impedance $Z_{n}(\omega)=R$, which we assume to be the same for both tunnel-junctions and leads.

We first consider the low bias regime $\mu\ll \Delta,\omega_{R}$, or equivalently small resistances $R$, with $\omega_{R}=1/RC$ being the bath 
frequency cut-off. We then obtain for the current $I_{1}$ for the tunneling of two spin-entangled electrons into separate  leads
\begin{equation}
\label{I1}
I_{1}=e\pi\mu\Gamma^{2}F_{d}^{2}(\delta r)\frac{e^{-4\gamma/g}}{\Gamma(4/g+2)}\left(\frac{2\mu}{\omega_{R}}\right)^{4/g}.
\end{equation}
In Eq.~(\ref{I1}) we introduced the  Gamma function $\Gamma(x)$ and the dimensionless tunnel-conductance $\Gamma= 4\pi\nu_{S}\nu_{l}|t_{0}|^{2}$ 
with $\nu_{S}$ and $\nu_{l}$ being the DOS per spin of the SC and the leads at the Fermi level $\mu_{S}$ and $\mu_{l}$, resp. Here 
$\gamma=0.5772$ is the Euler number. The exponent $4/g$ in Eq.~(\ref{I1}) with $g=R_{Q}/R$ is just two times the value  for single electron  
tunneling \cite{Devoret} via  one junction since the two tunneling events into different leads are uncorrelated.

In the large bias regime (and/or large resistances $R$) $\Delta\gg|\mu-E_{c}|\gg\omega_{R}$ we obtain 
\begin{equation}
I_{1}=e\pi\Gamma^{2}F_{d}^{2}(\delta r)\Theta(\mu-E_{c})(\mu-E_{c}),
\end{equation}
where small terms  $\sim e\pi\Gamma^{2}F_{d}^{2}(\delta r)\omega_{R}[{\cal O}(\omega_{R}/\mu)+{\cal O}(\omega_{R}/|\mu-E_{c}|)]$ have been 
neglected.
This shows a gap in $I_{1}$ for $\mu<E_{c}$ and $R\rightarrow \infty$ with $E_{c}=e^{2}/2C$ the charging energy which is a striking feature of 
the dynamical CB. 

We now turn to the case when two electrons coming from the same Cooper pair tunnel to the same lead 1 or 2 and first concentrate on the low bias 
case $\mu\ll \omega_{R}, \Delta$. When $\Delta\gg\omega_{R},E_{c}$ the process appears as a tunneling event of a charge $q=2e$ into the same lead 
with the result 
\begin{equation}
\label{2e}
I_{2}=e\pi\mu\Gamma^{2}\frac{e^{-8\gamma/g}}{\Gamma(8/g+2)}\left(\frac{2\mu}{\omega_{R}}\right)^{8/g}.
\end{equation}
The exponent $8/g$ shows that a dynamical CB effect due to a charge $q=2e$ is formed. The exponent of the power law decay in Eq.~(\ref{2e}) 
reacts quadratically with respect to the tunneling charge which is not surprising since the change of the junction capacitor's charging energy 
due to tunneling of a charge $q$ is $q^{2}/2C$. As a result, we obtain the ratio $I_{2}/I_{1}\propto (2\mu/\omega_{R})^{4/g}$. For values 
$\Delta\ll\omega_{R}$, e.g for small $R$, we obtain a similar result as in a Luttinger liquid, see Eq.~(\ref{currentI222})
\begin{equation}
\label{smallD}
I_{2}=e\pi\mu\Gamma^{2}A(g)\left(\frac{2\mu}{\omega_{R}}\right)^{4/g}\left(\frac{2\mu}{\Delta}\right)^{4/g}, 
\end{equation}
with $A(g)=(2e^{-\gamma})^{4/g}\Gamma^{4}(1/g+1/2)/\pi^{2}\Gamma(8/g+2)$.
Here the relative suppression of the current $I_{2}$ compared to $I_{1}$ is given essentially by $(2\mu/\Delta)^{4/g}$ and not by 
$(2\mu/\omega_{R})^{4/g}$ as in the case of an infinite $\Delta$.

In the large  voltage regime $\Delta,\mu\gg\omega_{R}$ we expect a Coulomb gap due to a charge $q=2e$. Indeed, in the parameter range 
$|\mu-2E_{c}|\gg\omega_{R}$ and $\Delta\gg|\mu-E_{c}|$ we obtain for $I_{2}$ again up to small contributions $\sim e\pi\Gamma^{2}\omega_{R}[{\cal 
O}(\omega_{R}/\mu)+{\cal O}(\omega_{R}/|\mu-2E_{c}|)]$
\begin{equation}
I_{2}=e\pi\Gamma^{2}\Theta(\mu-2E_{c})(\mu-2E_{c}).
\end{equation}
This shows that $I_{2}$ is small ($\propto \omega_{R}^{2}/|\mu-2E_{c}|$) in the regime $E_{c}<\mu<2E_{c}$, whereas $I_{1}$ is finite ($\propto 
F_{d}^{2}(\delta r)(\mu-E_{c})$).
\begin{figure}[ht]
\vspace{0.3cm}
\begin{center}
\hbox{\resizebox{5.1cm}{!}{\includegraphics{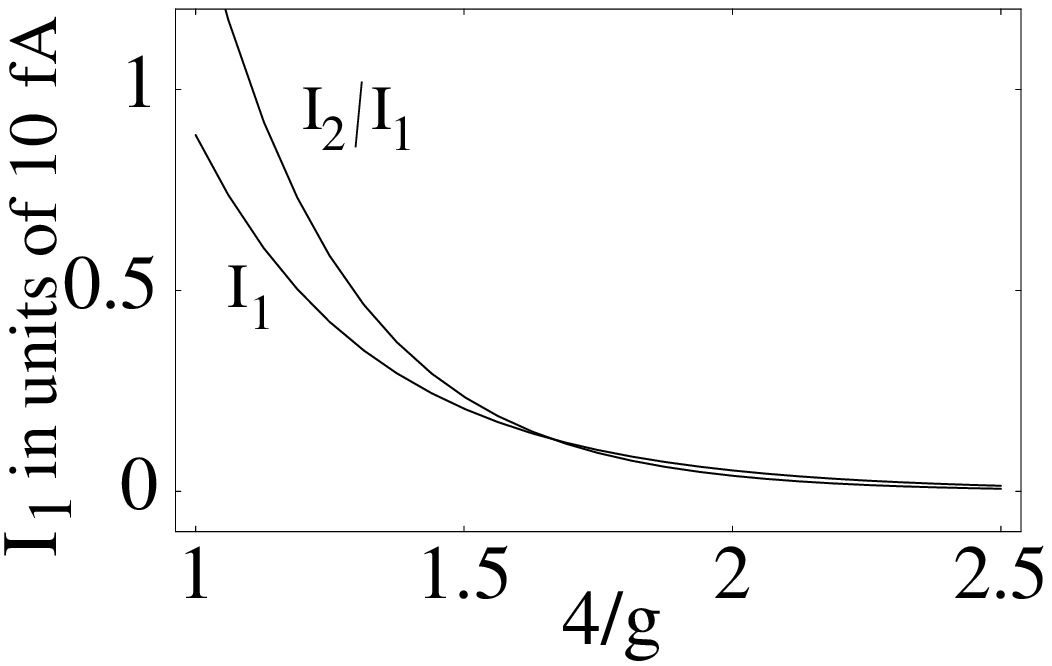}}\,\,\,\resizebox{4.8cm}{!}{\includegraphics{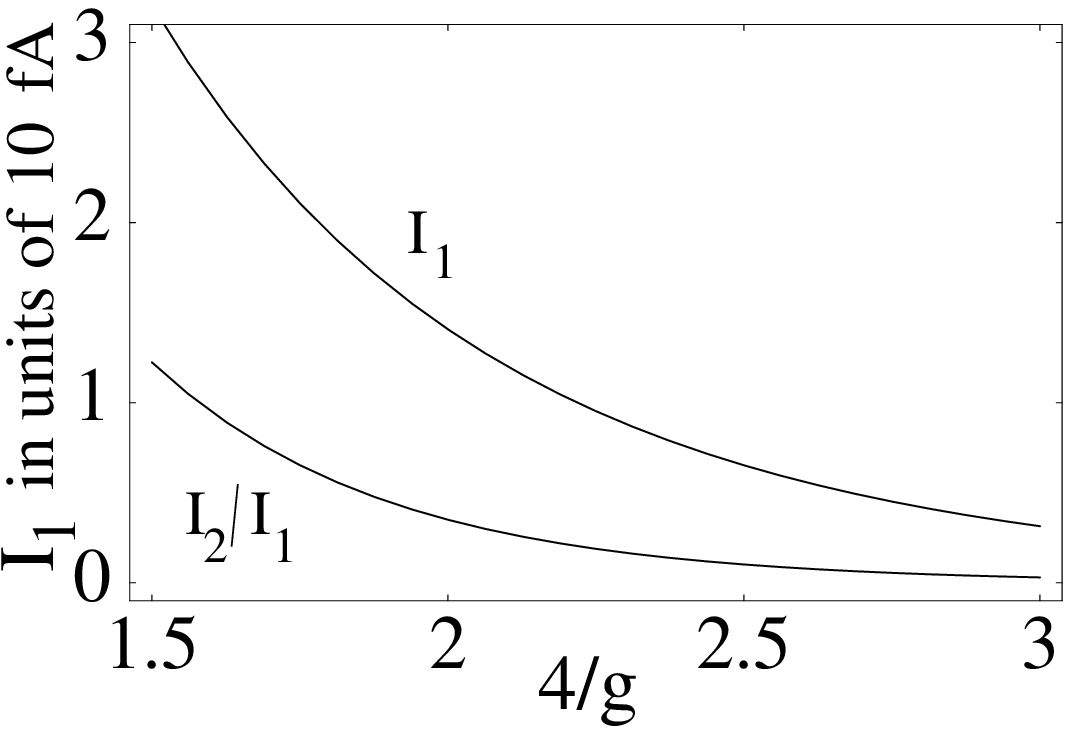}}}
\end{center}
\vspace{-0.3cm}
\caption{Current ratio $I_{2}/I_{1}$ (entangler efficiency) and current $I_{1}$ in the low bias regime, $\mu\ll\Delta,\omega_{R}$ and $\Delta\gg 
E_{c},\omega_{R}$, as a function of $4/g=4R/R_{Q}$. Chosen parameters: $E_{c}=0.1$ meV, $k_{F}\delta r=10$, $\Gamma=0.1$, and $\mu=5$ $\mu$eV 
(left plot), $\mu=15$ $\mu$eV (right plot). In the case of a 2D SC, $I_{1}$ and $I_{1}/I_{2}$ can be multiplied by 10. Adapted from \cite{RL2}.}
\label{plots}
\end{figure}
We now give numerical values for the current magnitudes and efficiencies of this entangler.
We first discuss the low bias regime $\mu\ll\Delta,\omega_{R}$. In Fig.~\ref{plots} we show the ratio $I_{2}/I_{1}$ (entangler efficiency) and  
$I_{1}$  for $\Delta\gg E_{c},\omega_{R}$ as a function of $4/g$ for realistic system parameters (see figure caption).
The plots show that a very efficient entangler can be expected for lead resistances $R\stackrel{<}{\sim}R_{Q}$. The total current is then on the 
order of $I_{1}\stackrel{>}{\sim}10$ fA.
In the large bias regime $\mu\gg\omega_{R}$ and for $E_{c}<\mu< 2E_{c}$ we obtain $I_{2}/I_{1}\propto (k_{F}\delta 
r)^{d-1}\omega_{R}^{2}/(2E_{c}-\mu)(\mu-E_{c})$, where we assume that $2E_{c}-\mu$ and $\mu-E_{c}\gg\omega_{R}$. For $\mu\simeq 1.5E_{c}$ and 
using $\omega_{R}=gE_{c}/\pi$ we obtain approximately $I_{2}/I_{1}\propto (k_{F}\delta r)^{d-1}g^{2}$. 
To have $I_{2}/I_{1}<1$ we demand that $g^{2}<0.01$ for $d=3$, and $g^{2}<0.1$ for $d=2$, $d$ being the effective dimension of the SC. Such small 
values of $g$ have been produced approximately in Cr leads \cite{Kuzmin}. For  $I_{1}$ we obtain $I_{1}\simeq e(k_{F}\delta 
r)^{1-d}(\mu-E_{c})\Gamma^{2}\simeq e(k_{F}\delta r)^{1-d}E_{c}\Gamma^{2}\simeq 2.5$ pA for $d=3$ and for the same parameters as  used in 
Fig.~\ref{plots}. 
This shows that $I_{1}$ is much larger than for low bias voltages, but an efficient entangler requires high lead resistances $R 
\stackrel{>}{\sim}10R_{Q}$.
Our discussion shows that it should be possible to implement the proposed device within state of the art techniques.

\section{Triple dot entangler}
\label{Tripledot}

\begin{figure}
{\centering \resizebox*{0.9\columnwidth}{!}{\includegraphics{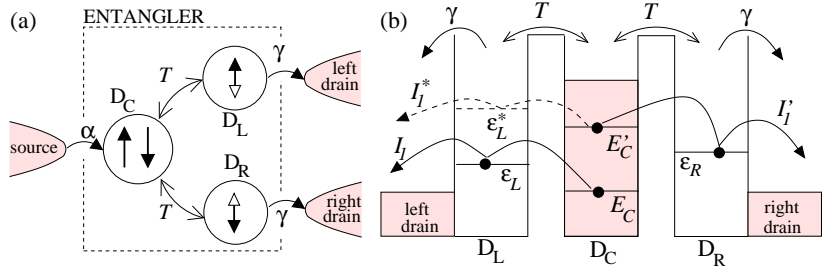}} \par}

\caption{\label{setup-triple}(a) Setup of the triple quantum dot entangler.
The central dot \protect\( D_{C}\protect \) has a singlet ground
state when \protect\( 2\protect \) electrons are present, and is
coupled coherently to the secondary dots \protect\( D_{L}\protect \)
and \protect\( D_{R}\protect \) with tunneling amplitudes \protect\( T\protect \).
The dots are each coupled incoherently to a different lead, with rate
\protect\( \alpha \protect \) and \protect\( \gamma \protect \).
(b) Energy level diagram for \emph{each} electron. The single-electron
currents \protect\( I_{1},I_{1}'\protect \) and \protect\( I_{1}^{*}\protect \)
are suppressed by the energy differences \protect\( \left| E_{C},E_{C}'-\epsilon _{L,R},\eps _{L}^{*}\right| \protect \),
while the simultaneous transport of the singlet pair from \protect\( D_{C}\protect \)
to \protect\( D_{L}\protect \) and \protect\( D_{R}\protect \) is
enhanced by the resonance \protect\( E_{C}+E_{C}'=\eps _{L}+\eps _{R}\protect \).
Adapted from \cite{Sar02}.}
\end{figure}

In this section we describe another scheme for the production of spin-entangled
electrons pairs based on a triple dot setup \cite{Sar02}. We shall use here an approach based on
perturbation theory that is quite transparent ---although it is less
rigorous than the master equation technique used in Ref. \cite{Sar02}.
The simple idea behind this entangler is described in Fig \ref{setup-triple}.
First we use the spin-singlet state occurring naturally in the ground
state of an asymmetric quantum dot \( D_{C} \) with an even number
of electrons \cite{Kik97}. Secondly, we use two additional quantum
dots \( D_{L,R} \) as energy filters; this provides an efficient
mechanism to enforce the simultaneous propagation of the singlet pair
into two separate drain leads --very much like the Andreev Entangler
discussed in Section~\ref{ssecAndreev}. An important point is that the
spin is conserved throughout the transport from \( D_{C} \) to the
drain lead (until the spin decoherence time is reached), so that we
only need to check the charge transport of the singlet state.

We assume that the chemical potential of the leads are arranged so
that only \( 0,1 \) or \( 2 \) (excess) electrons can occupy \( D_{C} \),
while \( 0 \) or \(1 \) electron can occupy \( D_{L,R} \). To simplify
notations, we assume a `symmetric' situation for the charging energy
in \( D_{C} \), namely that \( U_{C}(2)=U_{C}(0)\equiv 0 \), \( U_{C}(1)=-e^{2}/2C_{\Sigma }=:U \)
with \( U_{C}(N) \) the total Coulomb charging energy for \( N \)
excess electrons in \( D_{C} \), and \( C_{\Sigma } \) the total
capacitance of \( D_{C} \). It is crucial that \( D_{C} \) has an
even number of electrons when \( N=0 \) in order to have a singlet
ground state \( \ket {\up \down -\down \up } \) when \( N=2 \).
The total energies are \( E_{C}(0)\equiv 0,E_{C}(1)=\epsilon _{C}-U,\, \mathrm{and}\, E_{C}(2)=2\epsilon _{C} \),
where \( \epsilon _{C} \) is the lowest single-particle energy available
for the first excess electron. Therefore, the energy of the first
and second electron are \( E_{C}=\epsilon _{C}-U \) and \( E'_{C}=\epsilon _{C}+U \),
respectively. Similarly, we define \( E_{L,R}(0)\equiv 0 \) and \( E_{L,R}(1)=\epsilon _{L,R} \).
The transport is dominated by sequential tunneling, and we describe
the incoherent tunneling from the central source lead and \( D_{C} \)
by a tunneling rate \( \alpha  \), while \( \gamma  \) is the rate
of tunneling between \( D_{L} \) and \( D_{R} \) to their respective
drain leads. 

In the following, we shall show that it is possible to enhance by
resonance the simultaneous transport of the singlet across the triple-dot
structure, and to suppress single-electron transport carrying no entanglement.
In Ref. \cite{Sar02} we considered the quantum oscillations between
\( D_{C} \) and \( D_{L,R} \) ---described by the tunneling amplitude
\( T \)--- exactly, i.e., in infinite order. Here we shall restrict
ourselves to the lowest order. The first type of single-electron transport
\( I_{1} \), shown in Fig \ref{setup-triple}(b), corresponds to
the sequence

\begin{equation}
\label{trans-1}
I_{1}\, :\, \, \, 0\stackrel{\alpha }{\longrightarrow }C\left[ \stackrel{T}{\longleftrightarrow }L\stackrel{\gamma }{\longrightarrow }\right] 0\, 
\, \equiv \, \, 0\stackrel{\alpha }{\longrightarrow }C\stackrel{W_{1}}{\longrightarrow }0,
\end{equation}
where \( 0 \) denotes the situation with no excess electrons, \( C \)
denotes one electron in \( D_{C} \), and \( L \) one electron in
\( D_{L} \). (We do not describe here the situation obtained by replacing
\( L \) by \( R \)). In the right hand-side we have approximated
the coherent oscillations (\( T \)) and the incoherent tunneling
(\( \gamma  \)) to the drain leads (shown within the square brackets)
by a single rate \( W_{1} \). To find this rate, we consider that
the finite escape rate \( \gamma  \) to the drain leads broadens
the discrete energy level \( \eps _{L} \) in the secondary dot \( D_{L} \),
described by a Lorentzian density of state \begin{equation}
\label{lorent}
\rho _{L}(E)=\frac{1}{\pi }\frac{\gamma /2}{(E-\eps _{L})^{2}+(\gamma /2)^{2}}.
\end{equation}
Then \( W_{1} \) is given by the Fermi Golden rule
\begin{equation}
\label{w1}
W_{1}=2\pi |T|^{2}\rho _{L}(E_{C})=\frac{4\gamma T^{2}}{4\Delta _{1}^{2}+\gamma ^{2}}.
\end{equation}
 We have introduced the difference \( \Delta _{1}=E_{C}-\eps _{L}=\eps _{C}-U-\eps _{L} \)
between \( \eps _{L} \) and the energy \( E_{C} \) of the first
electron of the singlet state in \( D_{C} \). The diagram (\ref{trans-1})
corresponds to a 2-population problem, with the stationary current\begin{equation}
\label{cur1}
I_{1}=e\frac{\alpha W_{1}}{\alpha +W_{1}}=e\frac{4\alpha \gamma T^{2}}{\alpha (4\Delta _{1}^{2}+\gamma ^{2})+4\alpha T^{2}}.
\end{equation}

We can proceed similarly for the second process involving one-electron
transport:\begin{equation}
\label{trans-2}
I'_{1}\, :\, \, \, C\stackrel{\alpha }{\longrightarrow }CC\left[ \stackrel{T}{\longleftrightarrow }LC\stackrel{\gamma }{\longrightarrow }\right] 
C\equiv C\stackrel{\alpha }{\longrightarrow }CC\stackrel{W'_{1}}{\longrightarrow }C.
\end{equation}
The current \( I_{1}' \) is given by the same expression as Eq.~(\ref{cur1}),
with \( \Delta _{1} \) replaced by \( \Delta '_{1}=E'_{C}-\eps _{L}=\eps _{C}+U-\eps _{L} \)
(the difference between \( \eps _{L} \) and the energy \( E'_{C} \)
of the second electron of the singlet state \( CC \)). Therefore,
\( I_{1} \) and \( I_{1}' \) are suppressed by the energy differences
\( \Delta _{1} \) and \( \Delta _{1}' \). For simplicity, we now
take an almost symmetric setup \( \eps _{L}\simeq \eps _{R}\simeq \eps _{C}\then \Delta _{1}\simeq -\Delta _{1}'\simeq U \),
so that \( I_{1}\simeq I_{1}' \). 

The joint (simultaneous) transport of \( CC \) into \( LR \) propagates
the entanglement from \( D_{C} \) to the drain leads. We describe
it by the transition \begin{equation}
\label{transee}
I_{E}\, :\, \, \, 0\stackrel{\alpha }{\longrightarrow }C\stackrel{\alpha }{\longrightarrow }CC\left[ \stackrel{T}{\longleftrightarrow 
}LC\stackrel{T}{\longleftrightarrow }LR\stackrel{\gamma }{\longrightarrow }L\stackrel{\gamma }{\longrightarrow }\right] 0,
\end{equation}
 We approximate the double tunneling \( T \) and the escape to the
drain lead by a rate \( W_{E} \) given by a \( \second  \) order
Fermi Golden rule:\begin{equation}
\label{we}
W_{E}=\frac{2\gamma T^{4}}{\Delta _{E}^{2}+\gamma ^{2}}\frac{1}{\Delta _{1}^{'2}}
\end{equation}
with the two-particle energy difference \( \Delta _{E}=E_{CC}-E_{LR}=2\eps _{C}-\eps _{L}-\eps _{R}. \)
Note that \( W_{E} \) is also suppressed by \( \Delta _{1}' \),
which enters here as the energy difference between the initial state
\( CC \) and the virtual state \( LC \). We broadened the final
state \( LR \) with a rate \( 2\gamma  \) as the electrons can first
escape either to the left or to the right drain lead (i.e., \( LR\stackrel{\gamma }{\longrightarrow }\{L\, \, \mathrm{or}\, \, 
R\}\stackrel{\gamma }{\longrightarrow }0 \)).
Taking into account the additional channel involving the virtual state
\( CR \) (which gives approximately the same contribution \( W_{E} \)),
the transition diagram is\begin{equation}
\label{transE}
I_{E}\, :\, \, \, 0\stackrel{\alpha }{\longrightarrow }C\stackrel{\alpha }{\longrightarrow }CC\stackrel{2W_{E}}{\longrightarrow }0
\end{equation}
 and yields the stationary current\begin{equation}
\label{curE}
I_{E}=e\frac{2\alpha W_{E}}{\alpha +4W_{E}}=e\frac{4\gamma T^{4}}{\Delta _{1}^{'2}\left( \Delta _{E}^{2}+\gamma ^{2}\right) +8T^{4}\gamma /\alpha 
}.
\end{equation}
We now compare \( I_{E}/2 \) to the single electron currents \( I_{1} \)
and \( I_{1}' \). The entangler quality \( R \), defined by \begin{equation}
\label{qual}
\frac{I_{E}}{2I_{1}}>R,
\end{equation}
gives the ratio of the number of singlets to uncorrelated electrons
found in the drain leads. It yields the conditions \cite{Saragacomment}
\begin{equation}
\label{cond1}
T<U\sqrt{\frac{\alpha }{4\gamma R}},
\end{equation}
\begin{equation}
\label{cond2}
\gamma ,\Delta _{E}<\frac{T}{\sqrt{2R}},
\end{equation}
which correspond to Eqs.~(7) and~(8) of Ref. \cite{Sar02}. We note that
the sequence \begin{equation}
\label{11}
CC\stackrel{T}{\longleftrightarrow }LC\stackrel{T}{\longleftrightarrow }LR\stackrel{\gamma }{\longrightarrow }L\stackrel{\alpha }{\longrightarrow 
}LC,
\end{equation}
should not damage the entanglement shared by \( LR \). Indeed, the
transition \( LC\leftrightarrow CR \) is off-resonance, so that the
second electron in \( D_{L} \) quickly escapes to the lead before
it could tunnel back to \( D_{C} \) (which could create a new singlet
with the electron already present in \( D_{C} \) ). Finally, we find
that the current \( I_{E} \) saturates to \( e\alpha  \) if \( T^{4}\gg \alpha \gamma U^{2}/8 \).

So far we have assumed that no excited state could participate in
the transport. This is correct if the energy level spacings are large,
\( \delta \epsilon _{C,L,R}>2U \); in this case the single-electron
transport via the excited levels is suppressed even more than for \( I_{1} \)
and \( I_{1}' \). Unfortunately, this condition is not satisfied
in laterally-defined quantum dots, which are promising candidates
for an experimental implemention \cite{Wau95}. Below we estimate
the single-electron current \( I_{1}^{*} \) going via the excited
states of one given dot, e.g. \( D_{L} \). For simplicity we assume
the symmetric setup \( \eps _{C}=\eps _{L}=\eps _{R} \). 

We consider the excited state with the energy \( \eps _{L}^{*} \)
that is closest to the energy \( E_{C}' \) of the second electron
in the singlet state in \( D_{C} \). The non-entangled current \( I_{1}^{*} \)
is given by Eq.~(\ref{cur1}), with \( \Delta _{1} \) replaced by the
energy difference \( \Delta _{1}^{*}=E_{C}'-\eps ^{*}_{L} \). Introducing
the ratio \( R^{*}=I_{E}/2I_{1}^{*} \), we find the condition \( \Delta _{1}^{*}>\Delta _{1}\sqrt{R/R^{*}} \).
For sufficiently large \( R \) (e.g. \( R=100 \) as in \cite{Sar02}),
one can consider a reduced quality \( R^{*}\sim 10 \), which yields
\( \Delta _{1}^{*}>U/3 \). This corresponds to the minimal energy
difference found for a constant energy level spacing \( \delta \eps _{L}=2U/3 \)
\cite{Han03}. (In general, for odd \( N \) with \( \delta \eps _{L}=2U/N \)
we get \( \Delta _{1}^{*}=U/N \).) If \( \Delta _{1}^{*} \) is too
small, one should move the excited state away from the resonance by
increasing the energy level spacing \( \delta \eps _{L} \), which
for example could be achieved by applying a magnetic field \( B \)
perpendicular to the 2DEG, thus adding to the confinement energy the
Landau magnetic energy proportional to \( B \).

Finally, we comment on the validity of this perturbative approach.
It gives good results for the two-particle resonance defined by \( \Delta _{E}=0 \)
(where the entangled current dominates), as well as for the one-electron
resonance for \( I_{1}' \). However, it greatly overestimates
the resonance for \( I_{1} \) (at \( \Delta _{1}\simeq 0 \)), because
it naively neglects the two-electron channel by arguing that \( W_{E}\ll W_{1} \).
The rate for the single-electron loop is given by \( I_{1}/e\simeq \alpha \ll W_{1} \),
which allows the arrival of a second electron into \( D_{C} \) (with
rate \( \alpha  \)) and from there contributions from the two-electron
channel. One can consider a more complex Markovian chain including
both transition diagrams for \( I_{1} \) (Eq.~(\ref{trans-1})) and \( I_{E} \)
(Eq.~(\ref{transE})) -- however, in this case the approximation underestimates
the corresponding current. To obtain an accurate result for this resonance,
one must therefore follow the master equation approach used in Ref. \cite{Sar02}.

\section{Detection of spin-entanglement}
\label{Detection}

In this section we present schemes to measure the produced spin-entanglement in a transport experiment suitable for the above presented entangler 
devices.
One way to measure spin-singlet entangled states  is via shot noise experiments in a beamsplitter setup \cite{beamsplitter}. 

Another way to detect entanglement is to perform an experiment in which the Bell inequality~\cite{Bell} is violated.
\begin{figure}[h]
\centerline{\includegraphics[width=7cm]{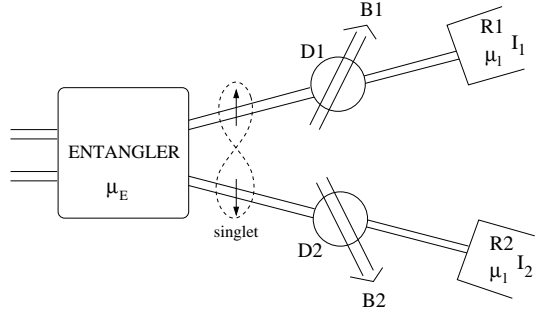}}
\vspace{1mm}
\caption{\small{The setup for measuring Bell inequalities: The entangler delivers a current of nonlocal singlet spin-pairs due to a bias voltage 
$\mu=\mu_{E}-\mu_{l}$. Subsequently, the two electrons in leads 1 and 2 pass two quantum dots D1 and D2, respectively,  which act as spin filters 
\cite{recher} so that only one spin direction, e.g. spin down, can pass the dots. The quantization axes for the spins are defined by the magnetic 
fields applied to the dots, which are in general different for D1 and D2. Since the quantum dot spin  filters act as a spin-to-charge converter, 
spin correlation measurements as required for measuring Bell inequalities can be reduced to measure current-current fluctuation correlators 
$\langle\delta I_{2}(t)\,\delta I_{1}(0)\rangle$ in reservoirs R1 and R2~\cite{Kawabata,Chtchelkatchev}.}}
\label{Bell}
\end{figure}
The Bell inequality describes correlations between spin-measurements of pairs of particles within the framework of a {\it local } theory. The 
Bell inequality measurement requires that a nonlocal entangled pair, e.g. a singlet, produced by the spin-entanglers, can be measured along three 
different, not mutually orthogonal, axes defined by unit vectors $\,\,{\hat{\bf a}}$, ${\hat{\bf b}}$ and ${\hat{\bf c}}$.
In a classical (local) theory the joint probabilities $P(i,j)$ satisfy the Bell inequality \cite{Sakurai}
\begin{equation}
\label{Bellinequality}
P({\hat{\bf a}}+, {\hat{\bf b}}+)\le P({\hat{\bf a}}+,{\hat{\bf c}}+) + P({\hat{\bf c}}+,{\hat{\bf b}}+).
\end{equation}
For example, $P({\hat{\bf a}}+, {\hat{\bf b}}+)$  is the probability that in a spin-correlation measurement between the spins in leads 1 and 2, 
see Fig.~\ref{Bell}, the measurement outcome for lead 1 is spin up when measured along the ${\hat{\bf a}}$-axis and the measurement in lead 2 
yields spin-up along the ${\hat{\bf b}}$-axis.
For a singlet state $|S\rangle=(|\!\uparrow\rangle_{1}|\downarrow\rangle_{2}-|\downarrow\rangle_{1}|\!\uparrow\rangle_{2})/\sqrt{2}$ the joint 
probability  $P({\hat{\bf a}}+, {\hat{\bf b}}+)$ becomes $ P({\hat{\bf a}}+, {\hat{\bf b}}+)=(1/2)\sin^{2}(\theta_{ab}/2)$. Here $1/2$ is just 
the probability to find particle 1 in the $\,\,{\hat{\bf a}},+$ state, and $\theta_{ab}$ denotes the angle between axis ${\hat{\bf a}}$ and 
${\hat{\bf b}}$. Similar results hold for the other functions in Eq.~(\ref{Bellinequality}). Therefore the Bell inequality for the singlet reads
\begin{equation}
\label{Bellinqual}
\sin^{2}\left(\frac{\theta_{ab}}{2}\right)\le\sin^{2}\left(\frac{\theta_{ac}}{2}\right)+\sin^{2}\left(\frac{\theta_{cb}}{2}\right).
\end{equation}
For a suitable choice of axes ${\hat{\bf a}}$, ${\hat{\bf b}}$ and ${\hat{\bf c}}$ and range of angles $\theta_{ij}$, this inequality is 
violated. For simplicity, we choose ${\hat{\bf a}}$, ${\hat{\bf b}}$ and ${\hat{\bf c}}$ to lie in a plane such that ${\hat{\bf c}}$ bisects the 
two directions defined by ${\hat{\bf a}}$ and ${\hat{\bf b}}$:
\begin{equation}
\theta_{ab}=2\theta,\,\,\,\,\theta_{ac}=\theta_{cb}\equiv\theta.
\end{equation}
The Bell inequality Eq.~(\ref{Bellinqual}) is then violated for 
\begin{equation}
0<\theta<\frac{\pi}{2}.
\end{equation}


To measure these joint probabilities $P(i,j)$ one can use spin filters in the outgoing leads of the entangler, see Fig.~\ref{Bell}. Such filters 
act as spin-to-charge converters where the spin information is transferred into the ability for the electrons to pass the filters. The (charge) 
current fluctuations of the measured electrons can then be detected in a reservoir placed after the spin filter. Such spin filters can be 
implemented by quantum dots in the spin filter regime as described theoretically in Ref. \cite{recher} and verified experimentally with a 
filtering efficiency up to 99.9 $\%$ \cite{Delftfilter}.
In the following we qualitatively describe the Bell measurement using the quantum dot spin filters.
We have seen that all we require are the probabilities $P(i,j)$. 
The directions  ${\hat{\bf a}}$, ${\hat{\bf b}}$ and ${\hat{\bf c}}$ are defined with a magnetic field along these axes \cite{fieldcomment} 
applied to the dots 1 and 2, see Fig.~\ref{Bell}. 
The principle of spin-measurement can be formulated in the following way.

Let us suppose we inject electrons into Fermi liquid leads via the entangler.
We have shown in the entangler setups containing quantum dots (Sections~\ref{ssecAndreev} and~\ref{Tripledot}) that pairs of electrons can be 
resonantly injected within some level width $\gamma_{e}$ around the energies $\varepsilon_{1,2}$ for leads 1,2. Therefore, the energies of 
electrons injected into leads  1,2 are well defined. In the case where the entangler is based on the Luttinger liquid 
(Section~\ref{sssluttinger}) or the finite resistance setup (Section~\ref{resistiveleads}), the injection is  not resonant and therefore not 
peaked in energy around some level $\varepsilon_{i}$ above the Fermi energy. In this case, the quantum dot spin filters in the leads themselves 
produce a resonance with width $\gamma_{D}$,  where $\gamma_{D}$ is the level broadening of the dot levels due to the coupling to the leads.

\begin{figure}[h]
\centerline{\includegraphics[width=7cm]{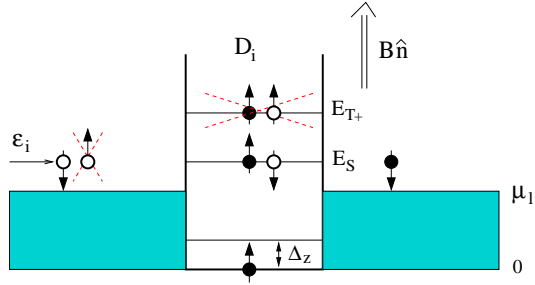}}
\vspace{1mm}
\caption{{\small The quantum dot as a spin to charge converter: The electrons are injected with energy $\varepsilon_{i}$ in lead $i=1,2$ above 
the chemical potential $\mu_{l}$. Since these energies can be tuned, with the entanglers or with the filters itself (see text),  such that 
$\varepsilon_{i}$= $E_{S}$ the transmission amplitude is very close to one if the spin is down (resonant transmission) with respect to direction 
$\hat n$ and strongly suppressed if the spin is up (cotunneling  process). Therefore the dot can act as a spin to charge converter where the spin 
information is converted to the possibility for the electron to pass the dot.}}
\label{Belldot}
\end{figure}
The quantum dot in lead $i=1,2$ is in the cotunneling  regime $(E_{S}-\mu_{l})>k_{B}T,\gamma_{D}$. Note that no voltage bias is applied to the 
quantum dot filters, see Fig. \ref{Belldot}. The quantum dot contains an odd number of electrons with a spin up ground state \cite{evenfilter}. 
The injected electron has energy $\varepsilon_{i}$ that coincides with the singlet energy $E_{S}$ (counted from $E_{\uparrow}=0$). The electron 
can now tunnel coherently  through the dot (resonant tunneling), but only if its spin is down. If the electron spin is up, it can only pass 
through the dot  via the virtual triplet state $|T_{+}\rangle$ which is strongly suppressed by energy conservation if 
$E_{T_{+}}-E_{S}>\gamma_{D}$ and $\gamma_{e}\le\gamma_{D}$. In addition, the Zeeman splitting $\Delta_{z}$ should be larger than $E_{S}-\mu_{l}$ 
in order to prevent excitations (spin down state, see Ref. \cite{recher}) on the dot induced by the tunnel-injected electron. Concisely, the 
regime of efficient spin-filtering is
\begin{equation}
\gamma_{e}\le\gamma_{D}<(E_{S}-\mu_{l}), E_{T_{+}}-E_{S},\,\, {\rm and}\,\,\,\,\,\, k_{B}T<(E_{S}-\mu_{l})<\Delta_{z}.
\end{equation}
In general, the incoming spin is in some  state $|\alpha\rangle$ and will not point along the quantization axis given by the magnetic field 
direction, i.e. $|\alpha\rangle=\lambda_{+}|\!\uparrow\rangle +\lambda_{-}|\!\downarrow\rangle$. This means that by measuring many electrons, all 
in the same state $|\alpha\rangle$, only  a fraction $|\lambda_{-}|^{2}$ will be in the down state and $|\lambda_{+}|^{2}=1-|\lambda_{-}|^{2}$ in 
the up state. To be specific: The probability that an electron passes through the filter is $|\lambda_{-}|^{2}$, provided that the transmission 
probability for a spin  down  electron is one (and zero for spin up), which is the case exactly at resonance $\varepsilon_{i}=E_{S}$ and for 
equal tunneling barriers on both sides of the dot~\cite{Datta}. So in principle, we have to repeat this experiment many times, i.e. with many 
singlets to get $|\lambda_{+}|^{2}$ or $|\lambda_{-}|^{2}$. But this is automatically provided by the entangler which exclusively delivers (pure) 
singlet states, one by one and such that there is a well defined (average) time between subsequent pairs which is much larger than the delay time 
within one pair (see previous sections). Therefore we can resolve single singlet pairs. 

How do we measure the successful passing of the electron through the dot?
The joint probability $P(i,j)$ quantifies correlations between spin measurements in lead 1 and 2 of the same entangled pair. Thus, this quantity 
should be directly related to the current-current fluctuation correlator $\int_{-\infty}^{+\infty}dt\,\langle \delta I_{2}(t)\,\delta 
I_{1}(0)\rangle$ measured in the reservoirs ${\rm R}_{1}$ and ${\rm R}_{2}$ if the filters are operated in the regime where only the spin 
direction to be measured can pass the dot. The current fluctuation operator in reservoir  $i=1,2$ is defined as $\delta I_{i}(t)=I_{i}(t)-\langle 
I_{i}\rangle$.  This quantitiy can be measured via the power spectrum of the shot noise $S(\omega)=\int_{-\infty}^{+\infty} dt\, e^{i\omega 
t}\,\langle \delta I_{2}(t)\,\delta I_{1}(0)\rangle$ at zero frequency $\omega$. Indeed, it was shown in Ref.~\cite{Samuelsson} that $P({\hat{\bf 
a}}\,\eta,{\hat{\bf b}}\,\eta')\propto S_{\eta,\eta'}({\hat{\bf a}},{\hat{\bf b}})$ where the zero frequency cross correlator is 
\begin{equation}
S_{\eta,\eta'}({\hat{\bf a}},{\hat{\bf b}})=2\int_{-\infty}^{+\infty}dt\,\langle \delta I_{\eta'{\hat{\bf b}}}(t)\delta I_{\eta{\hat{\bf 
a}}}(0)\rangle. 
\end{equation}
With $\eta$ and $\eta'$ we denote the spin directions ($\eta,\eta'=\uparrow,\downarrow$) with respect to the chosen axes ${\hat{\bf a}}$ and 
${\hat{\bf b}}$, respectively.
The proportionality factor between $P({\hat{\bf a}}\,\eta,{\hat{\bf b}}\,\eta')$  (the quantity of interest) and the cross correlator 
$S_{\eta,\eta'}({\hat{\bf a}},{\hat{\bf b}})$ can be eliminated by deviding $S_{\eta,\eta'}({\hat{\bf a}},{\hat{\bf b}})$ with 
$\sum_{\eta,\eta'}\,S_{\eta,\eta'}({\hat{\bf a}},{\hat{\bf b}})$~\cite{Samuelsson}. It was further pointed out in Ref.~\cite{Samuelsson} that the 
correlator $\langle \delta I_{\eta'{\hat{\bf b}}}(t)\delta I_{\eta{\hat{\bf a}}}(0)\rangle$ is only finite within the correlation time 
\cite{Samucomment} $\tau_{c}=1/\gamma_{e}$, i.e. for $|t|\stackrel{<}{\sim}\tau_{c}$. This sets some additional requirements to our entangler 
setups. The average time between subsequent arrivals of entangled pairs should be larger than this correlation time. This leads to the constraint
\begin{equation}
\label{correlationtime}
2e/I>1/\gamma_{e},
\end{equation}
where $I$ denotes the current of entangled pairs, i.e. the pair-split current calculated for various entangler systems in this review. The 
requirement Eq.~(\ref{correlationtime}) is always satisfied in our entanglers due to the weak tunneling regime. 

We conclude that the zero frequency correlator  $S_{\eta,\eta'}({\hat{\bf a}},{\hat{\bf b}})$ can be measured by a coincidence counting 
measurement of charges in the reservoirs ${\rm R}_{1}$ and ${\rm R}_{2}$ that collects statistics over a large number of pairs, all in the same 
singlet spin-state.

\section{Electron-holes entanglers without interaction}
\label{electronhole}

The entangler proposals presented in previous sections rely on entanglement sources which require interaction, e.g. Cooper pairs are paired up in 
singlet states due to an effective attractive interaction (mediated by phonons) between the electrons forming a Cooper pair. It was pointed out 
recently in Ref. \cite{Beenakker} and also in Refs. \cite{Samuelsson2,Lebedev,Lebedev2} that electronic entanglement could also be created 
without interaction.
In this section we would like to comment on two of the recent proposals
describing the production of electron-hole entanglement without interaction.
In the first one \cite{Beenakker}, quantum Hall edge states are used
 as 1D channels enabling the creation of entanglement via
the tunneling of one electron leaving a correlated hole behind it.
The entanglement is then dependent on how close the tunneling
amplitudes between different channels are; these
depend exponentially on the corresponding tunneling distances,
and are therefore different. Another problem 
lies in the random relative phases acquired by the final state
$\sum_{\sigma,\sigma'} \int d\epsilon \;
e^{i \phi_{\sigma,\sigma'}(\epsilon)} |t_{\sigma,\sigma'}(\epsilon)| 
 |\sigma,\sigma';\epsilon\rangle$
 after averaging over the energy bias, which could also degrade the entanglement as was pointed out in Ref.~\cite{dephasing}.

In the second one \cite{Lebedev}, it was shown that Bell inequalities
could be violated in a standard Y-junction at short times. The authors attribute
the related entanglement to the propagation of a singlet pair originating
from two electrons in the same orbital state. However, a short-time correlator can only probe single-electron properties as it takes a finite 
average time $~e/I$ to transfer two electrons; thus 2-electron singlets are not relevant (in accordance with \cite{Lebedev2}). Rather, the 
entanglement is shared by
electron-hole pairs across the two outgoing leads \( u \) and \( d \):
\(  \left| e,u,\uparrow \right\rangle \left| h,d,\uparrow \right\rangle +\left| e,u,\downarrow \right\rangle \left| h,d,\downarrow \right\rangle  
\),
very much like in Ref. \cite{Beenakker}. As the electron and the hole
are created by definition simultaneously, they are correlated at equal times and, as a result, the Bell inequality is only violated for short 
times \cite{Lebedev,Lebedev2}. On the other
hand, at longer times 2-electron correlations can appear in this setup.
Then, both singlets and triplets contribute, but the singlets from the {\it same} orbital state, $\epsilon=\epsilon'$, have measure zero in the 
correlator which involves a double integral over the energies $\epsilon,\epsilon'$ of the two electrons.

\section{Summary}
We have presented our theoretical work on the implementation of a solid state electron spin-entangler suitable for the subsequent detection of 
the spin-entanglement via charge transport measurements. In a superconductor, the source of spin-entanglement is provided by the spin-singlet 
nature of the Cooper pairs. Alternatively, a single quantum dot with a spin-singlet groundstate can be used. The transport channel for  the 
tunneling of two electrons of a pair into different normal leads is enhanced by exploiting Coulomb blockade effects between the two electrons. 
For this we proposed quantum dots, Luttinger liquids or resistive outgoing leads. We discussed a possible Bell-type measurement apparatus  based 
on quantum dots acting as spin filters.

\begin{acknowledgments}
This work was supported by the Swiss NSF, NCCR Basel, DARPA, and ARO. We thank C.W.J. Beenakker, G. Blatter, B. Coish, and V.N. Golovach for 
useful discussions.
\end{acknowledgments}


\newpage



\begin{chapthebibliography}{1}

\bibitem{Loss97}
D. Loss and D. P. DiVincenzo, Phys. Rev. A
\textbf{57}, 120 (1998), cond-mat/9701055.

\bibitem{Barenco}
A. Barenco, D. Deutsch, A. Ekert, and R. Josza, Phys. Rev. Lett. {\bf 74}, 4083 (1995).

\bibitem{Pfannkuche}
D. Pfannkuche, V. Gudmundsson, and P.A. Maksym, Phys. Rev. B {\bf 47}, 2244 (1993).

\bibitem{P1CBL}
M.-S. Choi, C. Bruder, and D. Loss, Phys. Rev. B {\bf 62}, 13569
(2000).

\bibitem{P3RSL}
P. Recher, E.V. Sukhorukov, and D. Loss, Phys. Rev. B {\bf 63},
165314 (2001).

\bibitem{P16Lesovik} G.B. Lesovik, T. Martin, and G. Blatter, Eur. Phys. J. B {\bf 24},
287 (2001).

\bibitem{P8RL}
P. Recher and D. Loss, Phys. Rev. B {\bf 65}, 165327 (2002).

\bibitem{P13BVBF}
C. Bena, S. Vishveshwara, L. Balents, and M.P.A. Fisher, Phys. \
Rev. \ Lett. {\bf 89}, 037901 (2002).

\bibitem{Samuelsson}
P. Samuelsson, E.V. Sukhorukov, and M. B\"uttiker, Phys. Rev. Lett. {\bf 91}, 157002 (2003).

\bibitem{RL2}
P. Recher and D. Loss, Phys. Rev. Lett. {\bf 91}, 267003 (2003).

\bibitem{Sar02} D.S. Saraga and D. Loss, Phys. Rev. Lett. {\bf 90}, 166803 (2003).

\bibitem{Oliver}
W.D. Oliver, F. Yamaguchi, and Y. Yamamoto, Phys. Rev. Lett. {\bf
88}, 037901 (2002).

\bibitem{Bos02}S. Bose and D. Home, Phys. Rev. Lett. \textbf{88}, 050401 (2002).

\bibitem{Saraga2}
D.S. Saraga, B.L. Altshuler, Daniel Loss, and R.M. Westervelt, Phys. Rev. Lett. {\bf 92}, 246803 (2004).

\bibitem{Kou97} L.P. Kouwenhoven, G. Sch\"on, and L.L. Sohn, Mesoscopic Electron
Transport, NATO ASI Series E: \ Applied Sciences-Vol.345, 1997,
Kluwer Academic Publishers, Amsterdam.

\bibitem{energyconservation}
This condition reflects energy conservation in the Andreev
tunneling event from the SC to the two QDs.

\bibitem{beamsplitter}
G. Burkard, D. Loss, and E.V. Sukhorukov, Phys.\ Rev.\ B {\bf
61}, R16 303 (2000). Subsequent considerations in this direction are discussed in Refs.\\
J.C. Egues, G. Burkard, and D. Loss,
Phys. Rev. Lett. {\bf 89}, 176401 (2002); G. Burkard and D. Loss,
Phys. Rev. Lett. {\bf 91}, 087903 (2003). For a comprehensive review on noise of spin-entangled electrons, see: J.C. Egues, P. Recher, D.S. 
Saraga, V.N. Golovach, G. Burkard, E.V. Sukhorukov, and D. Loss, in {\it Quantum Noise in Mesoscopic Physics}, pp 241-274, Kluwer, The 
Netherlands, 2003; cond-mat/0210498.

\bibitem{Merzbacher}E. Merzbacher, {\it Quantum Mechanics} 3rd ed., John
Wiley and Sons, New York, 1998, ch. 20.

\bibitem{cost}
This reduction factor of the current $I_{2}$ compared to the
resonant current $I_{1}$ reflects the energy cost in the virtual
states when two electrons tunnel via the same QD into the same
Fermi lead and are given by $U$ and/or $\Delta$. Since the
lifetime broadenings $\gamma_{1}$ and $\gamma_{2}$ of the two QDs
1 and 2 are small compared to $U$ and $\Delta$ such processes are
suppressed.

\bibitem{P4Klapwijk}
A.F. Volkov, P.H.C. Magnée, B.J. van Wees, and  T.M. Klapwijk,
Physica C {\bf 242}, 261 (1995).

\bibitem{Weiss}
J. Eroms, M. Tolkhien, D. Weiss, U. R\"ossler, J. De Boeck, and G. Borghs, Europhys. Lett. {\bf 58}, 569 (2002).

\bibitem{P5Bouchiat}
M. Kociak, A.Yu. Kasumov, S. Guéron, B. Reulet, I.I. Khodos, Yu.B.
Gorbatov, V.T. Volkov, L. Vaccarini, and H. Bouchiat, Phys. Rev.
Lett. {\bf 86}, 2416 (2001).

\bibitem{P6Bockrath}
M. Bockrath, D.H. Cobden, J. Lu, A.G. Rinzler, R.E. Smalley, L.Balents, and P.L. McEuen {\it Nature} {\bf 397}, 598 (1999).

\bibitem{P10Egger}
R. Egger and A. Gogolin, Phys. Rev. Lett. {\bf 79}, 5082 (1997);
R. Egger, Phys.\ Rev. \ Lett. {\bf 83}, 5547 (1999).

\bibitem{P11Kane}
C. Kane, L. Balents, and M.P.A.  Fisher, Phys. Rev. Lett. {\bf
79}, 5086 (1997).


\bibitem{Auslaender}
O.M. Auslaender, A. Yacoby, R. de Picciotto, K.W. Baldwin, L.N. Pfeiffer, and K.W. West, Phys. Rev. Lett. {\bf 84}, 1764 (2000).
 
\bibitem{P7rev}
For a review see e.g. H.J. Schulz, G. Cuniberti, and P. Pieri, in {\it Field Theories for Low-Dimensional Condensed Matter Systems}, G. Morandi 
{\it et al.} Eds. Springer, 2000; or J. von Delft and H. Schoeller, Annalen der
Physik, Vol. {\bf 4}, 225-305 (1998).

\bibitem{LLfootnote}
The interaction dependent constants $A_{b}$ are of order one for
not too strong interaction between electrons in the LL but are
decreasing when interaction in the LL-leads  is increased
\cite{P8RL}. Therefore in the case of substantially strong
interaction as it is present in metallic carbon nanotubes, the
pre-factors $A_{b}$ can help in addition to suppress $I_{2}$.

\bibitem{P21electronbunching}
Since $\gamma_{\rho -}>\gamma_{\rho +}$, it is more probable that
two electrons coming from the same Cooper pair travel in the same
direction than into different directions when injected into the
same LL-lead.


\bibitem{P12Balents}
L. Balents and R. Egger, Phys. Rev. B, {\bf 64} 035310 (2001).

\bibitem{Awschalom}  
J.M. Kikkawa and D.D. Awschalom, Phys. Rev. Lett. {\bf 80}, 4313 (1998).

\bibitem{InAs}
J. Nitta, T. Akazaki, H. Takayanagi, and K. Arai, Phys. Rev. B {\bf 46}, 14286 (1992); C. Nguyen, H. Kroemer, and E.L. Hu, Phys. Rev. Lett. {\bf 
69}, 2847 (1992).

\bibitem{Franceschi}
S. De Franceschi, F. Giazotto,  F. Beltram, L. Sorba, M. Lazzarino, and  A.
Franciosi, Appl. Phys. Lett. {\bf 73}, 3890 (1998).

\bibitem{Marsh}
A.M. Marsh, D.A. Williams, and H. Ahmed, Phys. Rev. B {\bf 50}, 8118 (1994).

\bibitem{Rimberg}
A.J. Rimberg, T.R. Ho, \c{C}. Kurdak, and J. Clarke, Phys. Rev. Lett. {\bf 78}, 2632 (1997).

\bibitem{Kuzmin}
L.S. Kuzmin,  Yu.V. Nazarov, D.B. Haviland, P. Delsing, and T. Claeson, Phys. Rev. Lett. {\bf 67}, 1161 (1991).

\bibitem{Devoret}
See e.g. G.-L. Ingold and Y.V. Nazarov, ch. 2 in H. Grabert and M.H. Devoret (eds.), Single Charge Tunneling, Plenum Press, New York, 1992.

\bibitem{oscillators}
Any lead impedance $Z_{n}(\omega)$  can be modeled with Eq.~(\ref{env}) via 
$Z^{-1}_{n}=\int_{-\infty}^{+\infty}dt\exp(-i\omega t)Y_{n}(t)$ with the admittance 
$Y_{n}(t)=\sum_{j=1}^{N}(\Theta(t)/L_{nj})\cos(t/\sqrt{L_{nj}C_{nj}})$.


\bibitem{Kik97}J. M. Kikkawa, I. P. Smorchkova, N. Samarth, and D. D. Awschalom,
Science \textbf{277}, 1284 (1997).
\bibitem{Saragacomment}
We also find the opposite conditions \( U\ll T\ll \gamma  \), which
are however incompatible with the perturbative approach.

\bibitem{Wau95}F. R. Waugh, M. J. Berry, D. J. Mar, and R. M. Westervelt, Phys. Rev.
Lett. \textbf{75}, 705 (1995); T. H. Oosterkamp, T. Fujisawa, W. G.
van der Wiel, K. Ishibashi, R. V. Hijman, S. Tarucha, and L. P. Kouwenhoven,
Nature \textbf{395}, 873 (1998).
\bibitem{Han03}Recent experiments have given the values \( \delta \eps =1.1\, \mathrm{meV},U=2.4\, \mathrm{meV} \);
see R. Hanson, B. Witkamp, L.M.K. Vandersypen, L.H. Willems van Beveren,
J.M. Elzerman, L.P. Kouwenhoven, cond-mat/0303139.
\bibitem{Bell}
J.S. Bell, Rev. Mod. Phys. {\bf 38}, 447 (1966).

\bibitem{recher}
P. Recher, E.V. Sukhorukov, and D. Loss, Phys. Rev. Lett. {\bf 85}, 1965 (2000).

\bibitem{Kawabata}
S. Kawabata, J. Phys. Soc. Jpn. {\bf 70}, 1210 (2001).

\bibitem{Chtchelkatchev}
N.M. Chtchelkatchev, G. Blatter, G. Lesovik, and T. Martin, Phys. Rev. B {\bf 66}, 161320(R) (2002).

\bibitem{Sakurai} J.J. Sakurai, {\em Modern Quantum Mechanics}, 
Addison Wesley, New York, 1985.

\bibitem{Delftfilter}
R. Hanson, L.M.K. Vandersypen, L.H. Willems van Beveren, J.M. Elzerman, I.T. Vink, and L.P. Kouwenhoven, cond-mat/0311414.

\bibitem{fieldcomment}
In our case, see Fig.~\ref{Belldot}, spin down is filtered by the quantum dots. Therefore, if we want to measure the spin e.g. along ${\hat{\bf 
a}}+$, we should apply the magnetic field in $-{\hat{\bf a}}$ direction.

\bibitem{evenfilter}
According to Ref.~\cite{recher}, an even number of electrons could also be considered for the spin filter.

\bibitem{Datta}
S. Datta, {\it Electronic Transport in Mesoscopic Systems}, Cambridge University Press, London, 1995.

\bibitem{Samucomment}
In Ref.~\cite{Samuelsson}, $\gamma_{e}$ is replaced by the voltage bias since no resonant injection, i.e. no quantum dots, are considered.

\bibitem{Beenakker}
C.W.J. Beenakker, C. Emary, M. Kindermann, and J.L. van Velsen, Phys. Rev. Lett. {\bf 91}, 147901 (2003).

\bibitem{Samuelsson2}
P. Samuelsson, E.V. Sukhorukov, and M. B\"uttiker, Phys. Rev. Lett. {\bf 92}, 026805 (2004).

\bibitem{Lebedev}
A.V. Lebedev, G.B. Lesovik, and G. Blatter, cond-mat/0311423.


\bibitem{Lebedev2}
A.V. Lebedev, G. Blatter, C.W.J. Beenakker, and G.B. Lesovik, Phys. Rev. B {\bf 69}, 235312, (2004).

\bibitem{dephasing}
J.L. van Velsen, M. Kindermann, C.W.J. Beenakker, Turk. J. Phys. {\bf 27}, 323 (2003), cond-mat/0307198.

\end{chapthebibliography}
\end{document}